\documentclass[a4paper,twoside]{article}

\usepackage{graphicx}
\usepackage{hhline,theorem,verbatim,euscript,url}
\usepackage[babel=true]{csquotes}
\usepackage{listings}
\usepackage{amssymb}
\usepackage{amstext}
\usepackage{mathtools}
\usepackage{textcomp}
\usepackage[ruled,vlined,linesnumbered]{algorithm2e}
\SetAlFnt{\footnotesize\sf}
\usepackage{subcaption}
\usepackage{float} 
\usepackage{color}
\usepackage{textcomp}
\usepackage{fancyvrb}
\usepackage{array, multirow,tabularx,csquotes,makecell}
\usepackage[export]{adjustbox}
\usepackage{calc}

\usepackage{multicol}
\usepackage{makecell}

\usepackage{pslatex}
\usepackage{apalike}
\newcolumntype{P}[1]{>{\raggedright\arraybackslash}p{#1}}

\usepackage[most]{tcolorbox}
\usepackage{lipsum}
\usepackage{enumitem}
\usepackage{hyperref}
\usepackage{booktabs}

\usepackage{cancel}

\tcbset{
	colback=gray!5,
	colframe=black,
	boxrule=0.5pt,
	arc=4pt,
	outer arc=4pt,
	boxsep=3pt,
	left=3pt,
	right=3pt,
	top=2pt,
	bottom=2pt,
}

\usepackage[most]{tcolorbox}
\usepackage{listings}
\usepackage{xcolor}

\lstdefinestyle{javaStyle}{
	language=Java,
	backgroundcolor=\color{gray!5},
	basicstyle=\ttfamily\scriptsize,
	stringstyle=\color{red!70!black},
	commentstyle=\color{gray!60}\itshape,
	breaklines=true,
	showstringspaces=false,
	tabsize=1
}

\newtheorem{theorem}{Theorem}
\newtheorem{definition}[theorem]{Definition}
\newtheorem{proposition}[theorem]{Proposition}

\usepackage{SCITEPRESS}

\makeatletter
\newcommand{\xnrightarrow}[2][]{%
	\mathrel{%
		\vphantom{\xrightarrow[#1]{#2}}%
		\ooalign{\hidewidth\neg@arrow\hidewidth\cr$\m@th\xrightarrow[#1]{#2}$\cr}%
	}%
}
\newcommand{\neg@arrow}{%
	$\m@th\vcenter{\hbox{%
			\rotatebox[origin=c]{-45}{\scalebox{1.5}[1]{$\m@th\scriptscriptstyle|$}}%
	}}$
}
\makeatother

\begin{document}
	\title{On the Soundness and Consistency of LLM Agents for Executing Test Cases Written in Natural Language}
	
\author{\authorname{S\'ebastien Salva\sup{1} and Redha Taguelmimt\sup{1}}
\affiliation{\sup{1}LIMOS - UMR CNRS 6158, Clermont Auvergne University, UCA, Aubière, France}
\affiliation{\sup{2}Department of Computing, Main University, MySecondTown, MyCountry}
	\email{sebastien.salva@uca.fr, redha.taguelmimt@uca.fr}
}

\abstract{
The use of natural language (NL) test cases for validating graphical user interface (GUI) applications is emerging as a promising direction to manually written executable test scripts, which are costly to develop and difficult to maintain. Recent advances in large language models (LLMs) have opened the possibility of the direct execution of NL test cases by LLM agents. This paper investigates this direction, focusing on the impact on NL test case unsoundness and on test case execution consistency. NL test cases are inherently unsound, as they may yield false failures due to ambiguous instructions or unpredictable agent behaviour. Furthermore, repeated executions of the same NL test case may lead to inconsistent outcomes, undermining test reliability. To address these challenges, we propose an algorithm for executing NL test cases with guardrail mechanisms and specialised agents that dynamically verify the correct execution of each test step. We introduce measures to evaluate the capabilities of LLMs in test execution and one measure to quantify execution consistency. We propose a definition of weak unsoundness to characterise contexts in which NL test case execution remains acceptable, with respect to the industrial quality levels Six Sigma. Our experimental evaluation with eight publicly available LLMs, ranging from 3B to 70B parameters,  demonstrates both the potential and current limitations of current LLM agents for GUI testing. Our experiments show that Meta Llama 3.1 70B demonstrates acceptable capabilities in NL test case execution with high execution consistency (above the level 3-sigma). We provide prototype tools, test suites, and results.
}
\keywords{GUI Application, Natural Language Test Case, Functional Testing, Soundness, Execution Consistency, LLM}

\onecolumn \maketitle \normalsize \vfill

\section{Introduction}

\begin{figure}[ht]
	\centering
	\scriptsize
	\begin{tcolorbox}[title=Scenario, width=0.49\textwidth]
		Open the website \href{https://www.uca.fr/en}{https://www.uca.fr/en}, 
		search for a page having news related to the \textbf{ARTEMIS} project.
	\end{tcolorbox}
	\caption{Scenario to search for ARTEMIS project news.}
	\label{fig:scenario}
\end{figure}

In industrial settings, the development and test of GUI applications often involves the use of scenarios to express user requirements, which are then refined into test cases written in natural language and subsequently into concrete test case scripts. As an illustrative example, Figure~\ref{fig:scenario} presents a simple scenario in which a user searches for pages related to a project on the website uca.fr/en. From this scenario, natural language (NL) test cases are derived; Figure~\ref{fig:tc} shows an example for the previous scenario. The NL test cases are then implemented as  executable test scripts, also shown in Figure~\ref{fig:seleniumtc}.

\begin{figure}[htbp]
	\centering
	\begin{tcolorbox}[title=Test Case, width=0.49\textwidth]
		\scriptsize
		\begin{itemize}[leftmargin=1em]
			\item Open the website \href{https://www.uca.fr/en}{https://www.uca.fr/en}
			\item Click on \texttt{European University}
			\item Click on \texttt{ALL NEWS}
			\item Assert that the page has links
			\item Assert that the page has links with the term \textbf{ARTEMIS}
		\end{itemize}
	\end{tcolorbox}
	\caption{Step-by-step test case to verify the presence of links containing the term ‘ARTEMIS’.}
	\label{fig:tc}
\end{figure}

\begin{figure}[htbp]
	\centering
	\begin{tcolorbox}[title=Selenium Test, width=0.49\textwidth]
		\vspace{-0.2cm}
		\begin{lstlisting}[style=javaStyle]
			@Test
			public void testSimple(){
				WebDriver driver = new FirefoxDriver();
				driver.get("https://www.uca.fr/en");
				WebElement element = driver.findElement(
				By.xpath("/html/body/header/div[3]/div[2]/nav/ul[1]/li[5]/a"));
				element.click();
				element = driver.findElement(
				By.xpath("/html/body/main/div[1]/div/div[1]/div[2]/div/div[3]/div/div/div/div[1]/a"));
				element.click();
				List<WebElement> links = driver.findElements(By.xpath("//a[@href]"));
				assertTrue(links.stream().anyMatch(
				l -> l.getText().contains("ARTEMIS")));
				driver.quit();}
		\end{lstlisting}
		\vspace{-0.2cm}
	\end{tcolorbox}
	\caption{Automated Selenium test implementing the test case of Figure \ref{fig:tc}.}
	\label{fig:seleniumtc}
\end{figure}

Writing concrete test cases such as the one in Figure \ref{fig:seleniumtc} is a time-consuming task, and the resulting source code can be difficult to maintain, particularly when UI elements change over time, e.g., the \texttt{"}/html/body/...\texttt{"} elements in the figure. As a consequence, functional tests are often neglected in practice. Artificial intelligence (AI) has the potential to reduce the effort involved in test development. Currently, two promising use cases warrant particular attention and evaluation:
1) the test case generation with Large Language Models (LLMs) from high-level scenarios or NL test cases. This approach can significantly reduce the manual effort required, although evaluating the correctness of the generated source code remains a challenging and time-consuming task \cite{SSTLLM24}; 2) the direct execution of NL test cases by means of agents powered by LLMs. Such LLM agents can indeed simulate user interactions with the GUI \cite{10638557,GPTDroid}, potentially allowing the complete execution of NL test cases.

This paper focuses on the second possibility by investigating the potential of LLM agents for testing GUI applications. 
Specifically, we address the following questions: can LLM  agents be effective for testing GUI applications with NL test cases? How does the use of these agents affect the test case soundness and the reproducibility of their execution (a.k.a. execution consistency)?

Without prolonging an unbearable suspense, it must be stated clearly that a NL test case is inherently unsound, i.e., it may reject an existing conformant implementation. The primary reasons for unsoundness are twofold: first, NL test cases may contain ambiguous actions whose interpretations may lead to a fail verdict; second, AI agents may behave unpredictably, or even hallucinate steps that are not specified. Additionally, the execution of NL test cases by AI agents introduces inconsistencies because different executions of the same test case may yield varying behaviours depending on factors such as prompt formulation or agent capabilities on performing navigation or on evaluating assertions. These inconsistencies undermine test case execution consistency. 

In this paper, we propose an algorithm and contexts under which NL test case unsoundness becomes acceptable. We formulate this notion of acceptability and define measures to evaluate test case execution consistency. Specifically, we propose an algorithm for executing NL test cases with guardrail mechanisms. It dynamically completes the test steps and controls them, enabling evaluation of whether each step is executed as intended. We leverage a set of specialised agents to handle specific actions: navigation through the GUI, readiness of the GUI to perform the next step, and evaluation of assertions. 

To assess the capabilities of these agents in performing these actions, we define three measures. In addition, we introduce another measure to quantify the execution consistency of NL test cases---that is, the ability of our algorithm to produce stable and repeatable test outcomes.

We also study the NL test case unsoundness and introduce a definition of weak unsoundness, which characterises acceptable contexts in which NL test cases can be reliably executed by LLM agents.

Furthermore, this paper presents experiments evaluating the capabilities of eight publicly available LLMs, which are deployable on local servers, in executing NL test cases. 
Our results show that while some LLMs are effective, further work is required to enhance their capabilities.

In summary, the major contributions of this paper are:

\begin{itemize}
\item an algorithm for executing NL test cases, aimed at verifying the correct execution of each test step;
\item three measures for evaluating the capabilities of LLMs in test execution, and one measure for assessing the test case execution consistency;
\item the definition of weak unsoundness of NL test cases, along with propositions characterising realistic executions;
\item two prototype tools and corresponding test suites, made publicly available in \cite{companion25}. 
\end{itemize}

The paper is organised as follows: Section~\ref{sec:relatedwork} reviews related work and outlines our motivations. Section~\ref{sec:tcexec} presents the algorithm for executing NL test cases. Sections~\ref{sec:consistency} and~\ref{sec:soundness} discuss test case execution consistency and the notion of unsoundness, respectively. Experimental results are reported in Section~\ref{sec:evaluation}. Finally, Section~\ref{sec:conclusion} concludes the paper and outlines directions for future work.

\section{Related work}
	\label{sec:relatedwork}

AI has been applied to various software engineering activities, with steadily increasing adoption over the past decade. Some surveys covered the use of machine learning \cite{Allamanis18} or LLMs for software engineering \cite{LLM4SE}. Several AI-based approaches have been proposed for program repair \cite{fan2023automated,jiang2023impact,Yasunaga20,8827954,didact}, source code generation \cite{pmlr-v119-kanade20a,austin2021program,Helander6862,SalvaS25} or to audit the security of applications \cite{ma2024combiningfinetuningllmbasedagents}.
	
AI and more specifically LLM-based approaches have also been proposed for testing. 		
ChatUniTest \cite{ChatUniTest} is an LLM-based automated unit test generation framework. It incorporates an adaptive focal context mechanism to encompass valuable context in prompts and adheres to a generation-validation-repair mechanism to rectify errors in generated unit tests. The evaluation shows that ChatUniTest outperforms TestSpark and EvoSuite in half of the evaluated projects, achieving the highest overall line coverage.

Fuzz testing, which aims at providing a program with unexpected or random inputs to identify bugs, is also currently improved with AI. For example, LLMFuzz \cite{LLMFuzz} is a testing tool that calls LLMs to generate inputs for fuzzing Deep Learning libraries.

Closer to our work, a few algorithms that extend Android crawlers were recently proposed to detect bugs. Instead of using meta-heuristics \cite{salva:hal-02019705}, they call LLM or LLM agents to better cover a mobile application. DroidAgent \cite{10638557} is a multi-agent tool for autonomous test generation for Android applications. It gathers three agents to interact, observe, and evaluate the success of tasks. To optimise the application coverage, the agents have three types of memory: short-term memory (GUI state), long-term memory (history of performed tasks) and spatial memory to record observations made after interacting with specific widgets. The study shows that DroidAgent outperforms existing state-of-the-art tools in terms of coverage, achieving 10\% higher coverage than the baseline. 
GPTDroid \cite{GPTDroid} is another tool for Android testing that covers an application to detect bugs. It extracts the GUI context, uses a mechanism that records past testing interactions (activities, widgets, operations), encodes them into prompt questions for the LLM, decodes the LLM answer into actionable operations to execute on the application, and iterates the whole process. Results on 93 popular applications demonstrate that GPTDroid achieves 32\% higher coverage than the baseline. 
These approaches dynamically explore the application’s navigation tree using LLM agents to find bugs. Since this process operates without predefined test cases, it lacks repeatability, making consistent test execution infeasible. In contrast, our work focuses on the execution of conformance test cases and on test case execution consistency.

The exploratory study presented in \cite{SSTLLM24} investigates how LLMs can support system test development---that is, the generation of NL test cases  as well as the creation of concrete test cases through specialised LLM prompts. Their study, conducted on one subject, shows that when user requirements are fully given, LLMs are able to provide NL test cases that cover most of them. However, generating concrete test cases is more challenging, as these require at least 30\% of modifications before becoming executable. In contrast, our work assumes the availability of NL test cases and investigates the alternative possibility of executing them directly.

\section{Test Case Execution}
\label{sec:tcexec}
We assume having a GUI application under test, denoted $AUT$, from which we can extract the DOM structure or screenshots can be captured. Furthermore, the GUI allows user interaction. 

We consider having a set of NL test cases, where a NL test case is represented as an ordered sequence of navigation actions (e.g., Click on ‘Sign in’) followed by a sequence of assertions (e.g., Assert that an element x is present). Formally, a NL test case is denoted as $tc = a_1 \dots a_k \, \mathcal{A}_{k+1} \dots \mathcal{A}_l$, where $a_1 \dots a_k$ are navigation actions  and $\mathcal{A}_{k+1} \dots \mathcal{A}_l$ are assertions. Each assertion $\mathcal{A}$ may be a simple predicate or a composite expression formed through conjunctions and disjunctions of simpler assertions.

\begin{figure*}[!ht]
	\centering
	\begin{subfigure}{0.49\textwidth}
		\includegraphics[width=.5\textwidth]{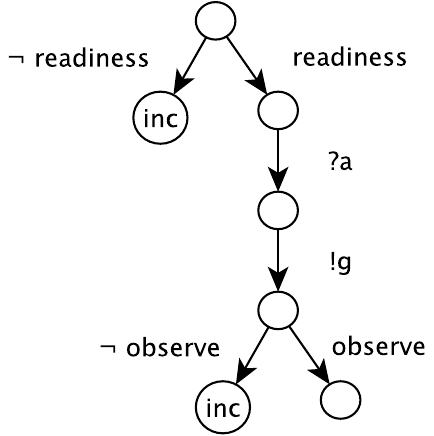}
		\caption{}
		\label{fig:schema1}
	\end{subfigure}
	\begin{subfigure}{0.49\textwidth}
		\includegraphics[width=.6\textwidth]{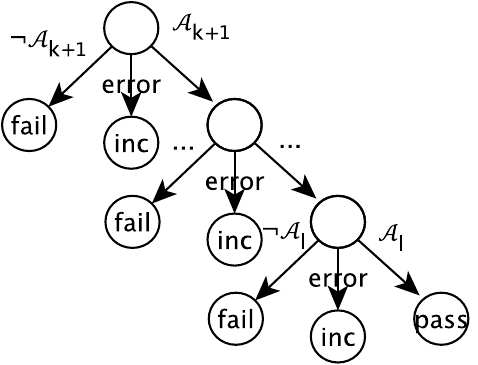}
		\caption{}	
		\label{fig:schema2}
	\end{subfigure}
	\caption{Illustration of the test case execution algorithm. (a) Navigation actions are injected with internal actions, readiness and observe. (b) Assetions are incrementally evaluated to determine the final verdict.}
	\label{fig:schema}
\end{figure*}

The experimentation of $AUT$ with a NL test case $tc$ using LLM agents introduces several challenges, including the soundness of the test case and the consistency of the resulting verdicts. Another difficulty concerns observability. When executing a subsequence of navigation actions from $tc$, the agents behaviours do not allow to observe how the $AUT$ responds. If a wrong GUI is returned by $AUT$ after an interaction, an assertion may fail later, but it remains difficult to identify which specific interaction caused the fault. 

With this in mind, we propose an algorithm that supplements on the fly a NL test case with additional internal guardrail actions. These actions aim at orchestrating the execution of agents and supervise how the $AUT$ behaves, ensuring that every navigation action is both feasible and carried out as expected. Specifically, we introduce two internal actions, denoted $readiness$ and $observe$, which are systematically injected for every navigation action, as illustrated in Figure~\ref{fig:schema1}. These actions are said to be internal in the sense that they are not observable during execution. $readiness$ checks whether the next navigation action is executable, i.e., whether the necessary UI elements are present on the interface to perform the action. In turn, $observe$ checks whether the last executed navigation action has led to a visible change in the GUI and whether the agent has successfully completed it. Both internal actions also contribute to evaluating the consistency and reliability of the test case execution (Section \ref{sec:consistency}). In addition to these internal actions, our algorithm adds to the test case the notion of inputs, i.e. navigation actions, starting by the symbol "?" applied or sent to $AUT$, and outputs, i.e. GUIs initiated by $AUT$, starting by the symbol "!" (see Figure~\ref{fig:schema1}).

\begin{algorithm}[htbp]
	\begin{footnotesize}
\SetKwInOut{Input}{input} \SetKwInOut{Output}{output}
		
\Input{Test case $tc=a_1 \dots a_k \, \mathcal{A}_{k+1} \dots \mathcal{A}_l$}
\Output{Test verdict, IOLTS $tc|AUT=<Q,L\cup \T,\rightarrow,q_0>$}

\ForEach{$a_i \in tc$}{
Add $q_{i-1} \xrightarrow{?a_i} q_{i,1} \xrightarrow{!g_i} q_{i,2} \xrightarrow{observe} q_{i,3}  \xrightarrow{readiness} q_{i}$, $q_{i,2} \xrightarrow{\neg observe} INC$, $q_{i,3}  \xrightarrow{\neg readiness} INC$ to $\rightarrow$\;
Execute the actions $a_i$, $observe$, $readiness$ and Cover $tc|AUT$\; 
}		

\ForEach{$\mathcal{A}_{i} \in tc$}{
\If {$i \neq l$} {
Add $q_{i-1} \xrightarrow{\mathcal{A}_{i}} q_{i}$ to $\rightarrow$
}
\Else {
Add $q_{i-1} \xrightarrow{\mathcal{A}_{i}} PASS$ to $\rightarrow$\;}

Add $q_{i-1} \xrightarrow{\neg \mathcal{A}_{i}} FAIL$ to $\rightarrow$\;
Add $q_{i-1} \xrightarrow{error} INC$ to $\rightarrow$\;

Execute $\mathcal{A}_{i}$ and cover $tc|AUT$\;
}
 
	\end{footnotesize}
	\caption{NL test case execution}
	\label{algo:A1}
\end{algorithm}

All this is reflected in Algorithm \ref{algo:A1}. It incrementally unfolds the NL test case $tc$ into a tree structure expressing the synchronous execution of $tc$ on $AUT$. We propose to model this execution by using an Input Output Labelled Transition System (IOLTS) denoted $tc|AUT$ whose edges are labelled with actions and terminal nodes are labelled with test verdicts, $tc|AUT=<Q, L \cup \T,\rightarrow,q_0>$, with $Q$ the state set, $L=L_I\cup L_O$ the set of inputs and outputs, respectively, $\T$ the set of internal actions, $\rightarrow$ the transition set and $q_0$ the initial state. 

Algorithm \ref{algo:A1} starts by covering every navigation action $a_i$ in $a_1 \dots a_k$. It encapsulates the action $a_i$ flanked by its corresponding actions $readiness (a_i)$ and $observe$, as illustrated in Figure~\ref{fig:schema1}. If either $readiness (a_i)$ or $observe$ are evaluated to false, the test case cannot be executed because the action cannot be performed or has not updated the GUI, and an Inconclusive verdict is returned. Otherwise, the algorithm continues with the next action. When the algorithm reaches the assertions $\mathcal{A}_{k+1} \dots \mathcal{A}_l$, it incrementally evaluates them, as illustrated in Figure \ref{fig:schema2}. If an assertion evaluates to false, then a Fail verdict is returned. The Inconclusive verdict is also reached in case an error is triggered by an agent. Otherwise, a Pass verdict is given. 

We now detail how every action (navigation action, assertion, readiness, observe) are performed:

\subsection{Navigation Action Execution}

Navigation actions are carried out by an agent that integrates a LLM for reasoning and decision-making, a memory module, 
and a function invocation interface. 

The memory module allows to store the current execution state (finished, error), step queues, intermediate results, and also incorporates an historical context, which potentially allows the agent to revisit previously accessed pages and make informed decisions based on past interactions. 
The function invocation interface enables the agent to interact programmatically with web content through the Document Object Model (DOM) or to interpret visual information extracted from screenshots. 

The agent is invoked using prompts we wrote and structured following the patterns proposed in \cite{prompt-pattern-cat}. We here considered: 1. the ``Fact Checklist" pattern that asks  the agent to output a list of facts included in the final answer, thereby enhancing transparency on the results provided; 2. the ``Template" pattern that enforces an output format in JSON, which includes fields such as extracted facts, results, and a boolean indicator of task completion status; 3. The ``Recipe" pattern, which builds a complete sequence of steps given some partially “ingredients” to achieve the given task. The prompt is given in Annex in Figure \ref{fig:navprompt}.

\subsection{Observe and Readiness Actions Execution}

The $observe$ action checks whether the last navigation action was executed successfully and whether it resulted in an update to the GUI (i.e., whether a detectable change occurred). This action returns both the content of the currently displayed page and a boolean value indicating the success of the agent’s navigation attempt and the consequent interface update. This action is executed without issuing a new command to an agent.
 
$readiness(a)$ checks whether the current GUI is ready for the execution of the upcoming navigation action $a$, i.e., whether the requisite UI elements are present. $readiness(a)$ is defined by the predicate logic formula $readiness\_strict(a) \vee (\neg readiness\_strict(a) \implies readiness\_w\_agent(a))$, which evaluates to a boolean value. 
Here, $readiness\_strict$ is a boolean procedure that performs the following steps: 1. translation of $a$ into a predicate; 2. evaluation of the predicate with the formula $\bigwedge_1^k \phi_i$, where $\phi_i$ are predicates that express actions performed on the GUI and constraints. For instance, consider the navigation action ``Click the link `Sign in'". A sub-formula $\phi$ written to evaluate whether this action can be performed on the GUI can be: $\forall x, \text{Click}(x) \implies \text{In}(x, content)$  with the predicates $\text{Click}(x) \implies \text{``Element x is clicked"}$ and $\text{In}(x) \implies \text{``Element x is in the GUI content"}$. Table \ref{table:readiness} lists the sub-formula we designed to perform the evaluation of $readiness\_strict$. 

\begin{table}[htbp]
	\centering
		\caption{Examples of formulas for evaluating GUI readiness.}
	\begin{tabular}{|p{1.5cm}|p{5cm}|}
		\hline
		\textbf{Action} & \textbf{Formula} \\
		\hline
		click(x) & $\forall x, \text{Click}(x) \implies \text{In}(x, \text{content}) \land (\text{Type}(x, \text{"link"}) \lor (\text{Type}(x, \text{"button"}))$ \\
		\hline
		select(x, v) & $\forall x, \text{Select}(x, v) \implies  \text{In}(x, \text{content}) \land \text{Type}(x, \text{"list"}) \land \text{In}(v, \text{option(x)})$ \\
		\hline
		check(x) & $\forall x, \text{Check}(x) \implies \text{In}(x, \text{content}) \land \text{Type}(x, \text{"checkbox"}) \land \lnot \text{Checked}(x)$ \\
		\hline
		uncheck(x) & $\forall x, \text{Uncheck}(x) \implies \text{In}(x, \text{content}) \land \text{Type}(x, \text{"checkbox"}) \land \text{Checked}(x)$ \\
		\hline
		fill(x, v) & $\forall x, \text{Fill}(x, v) \implies In(x,\text{content}) \land \text{Type}(x, \text{"input"})$ \\
		\hline
	\end{tabular}
	\label{table:readiness}
\end{table}

Using $readiness\_strict(a)$ offers the advantage of providing an unambiguous guarantee that the subsequent action can be executed directly on the GUI. However, writing a fully exhaustive formula is inherently challenging, given the dynamic and evolving nature of GUI interactions. 
This is why we use the second boolean procedure, $readiness\_w\_agent(a)$. An agent is here requested to assess GUI readiness by extracting all the UI elements of the GUI and by checking if the action $a$ can be executed with these elements. We request the agent with a prompt we structured with the following patterns \cite{prompt-pattern-cat}: ``Fact Check List" and ``Template", similar to the prompt written to perform navigation actions. We also employed the ``Chain of Thought" prompt engineering technique \cite{CoT22} to enable the reasoning capabilities of the agent by producing intermediate steps. The prompt is given in Annex in Figure \ref{fig:evalprompt}.

\subsection{Assertion Evaluation}

The most reliable way to ensure the validity of an assertion $\mathcal{A}$ is to express it using a strict logical formula. However, automatically translating textual assertions into such formal representations is often a long and non-trivial activity. To address this, we adopt a strategy analogous to the one used with $readiness$. An assertion $\mathcal{A}$ is evaluated by the logical formula $assert\_strict(\mathcal{A}) \vee (\neg assert\_strict(\mathcal{A}) \implies assert\_w\_agent(\mathcal{A}))$. Here, $assert\_strict$ refers to a boolean procedure that performs the following steps: 1. split of $\mathcal{A}$ into individual elementary assertions separated by operators; 2. translation of the elementary assertions into predicates; 3. evaluation of every predicate by means of the formula  $\bigwedge_{i=1}^k \phi_i$; 4. evaluation of the final boolean according to the operators in $\mathcal{A}$. The sub-formula $\phi_i$ expresses simple but usual generic assertions considered in GUI testing. For example,  ``Assert that 'x' is present" is express by $\forall x, Present(x) \implies In(x,page)$. Table \ref{table:assert} lists several sub-formula. 

\begin{table}[htbp]
	\centering
	\caption{Examples of assertions expressed with formulas.}
	\begin{tabular}{|p{1.5cm}|p{5cm}|}
		\hline
		\textbf{Assertion} & \textbf{Formula} \\
		\hline
		 'x' is present & $\forall x, \text{IsPresent}(x) \implies \text{In}(x, \text{content})$\\
		 \hline
		 'x' is not present & $\forall x, \lnot \text{IsPresent}(x) \implies \lnot \text{In}(x, \text{content})$\\
		 \hline
		 'x' is checked & $\forall x, \text{IsChecked}(x) \implies \text{In}(x, \text{content}) \land  \text{Checked}(x)$\\
		 \hline
		 'x' is visible & $\forall x, \text{IsVisible}(x) \implies \text{In}(x, \text{content}) \land  \text{Visible}(x)$\\
		 \hline
	\end{tabular}
		\label{table:assert}
\end{table}

The predicate $assert\_w\_agent(\mathcal{A})$ expresses the invocation of an agent to evaluate the assertion. This allows the use of more complex, but potentially more ambiguous, assertions such as ``Assert that the page has links". The agent is requested with another prompt also constructed with the patterns ``Fact Check List", ``Template" and ``Recipe". The latter is specifically used to force the agent
to focus on the UI elements relevant to the assertion, before evaluating them and giving a final verdict. It also uses ``Chain of Thought" to explicitly require the generation of intermediate steps before generating the final response. The prompt is given in Annex in Figure \ref{fig:assertprompt}.

\section{Test Case Execution Consistency}
\label{sec:consistency}

We now propose a measure to assess the estimated consistency of test case execution w.r.t. the chosen LLM agents used to evaluate readiness, assertions, or to interact with GUIs. 
We hence consider having the following agents: $agent_{nav}$ for performing navigation, $agent_{readiness}$ for evaluating the readiness of the GUI, and $agent_{assert}$ for evaluating assertions.

Given a test case $tc$, we want to estimate how stable the test verdicts of $tc$ should be across multiple potential runs. 
A high consistency score indicates the test passes or fails in a predictable way, while a low score suggests inconsistent behaviour. If the score is low, the test should not be executed with the current set of agents. 

A naive way to measure consistency would be to run a test case multiple times and check whether the verdicts and observed behaviours remain unchanged across runs. This approach is time consuming. Instead, we exploit the fact that the consistency of the test case execution depends on the abilities of each agent to perform its tasks in a predictable way. Hence, we initially assess the ability of every agent to perform its tasks by computing the standard deviation on its binary results 
, which measures the variation of the values of a variable. 
Let $p$ denote the probability that a given agent successfully performs its tasks. 
The standard deviation of the results produced by an agent is given by $0 \leq \sigma(agent)=\sqrt{p(1-p)}\leq 1$, with $p$ the probability of success for doing its tasks. The closer $\sigma(agent)$ is to 0, the more stable its execution is across multiple runs. To compute $\sigma(agent)$, we use a predefined set of test cases in which the expected outcomes (success/failure) of actions are known in advance. Each specialised agent is evaluated on this set, and its observed successes and failures provide $p$. This procedure is described in greater detail in Section \ref{sec:evaluation} in relation to RQ1.

We recall that a test case is made up of four kinds of actions: readiness, navigation, assertion, and observation. Among these, three are performed by agents---$readiness$, navigation actions, and assertions. To measure the consistency of their executions, we define three measures as follows:

First, $s_r(a)$ measures the consistency of $readiness(a)$, with $a$ a navigation action. Recall that the action $readiness(a)$ is either performed by $readiness\_strict(a)$ or $readiness\_w\_agent(a)$. $readiness$ $\_strict$ is expressed by a formula and is consistent. When the formula evaluates to false, we use $readiness\_w\_agent(a)$, which itself calls $agent_{readiness}$. The execution consistency of $readiness\_w$ $\_agent(a)$ is related to the standard deviation of $agent_{readiness}$. $s_r$ is hence defined as: 
$$0\leq s_r(a) \leq 1 =_{def} \left\{
\begin{array}{l}
	1 \text{ if } readiness\_strict(a)=1\\
	1-2\sigma(agent_{readiness}) \text{ otherwise }
\end{array}
\right.$$

Second, $s_n(a)$ measures the consistency of performing a navigation action $a$. Unlike $readiness$, this execution is exclusively made by $agent_{nav}$. Consequently, $s_n$ is defined as: $$0\leq s_n(a) = 1-2\sigma(agent_{nav})\leq 1 $$
	
Third, $s_a(\mathcal{A})$ measures the consistency of the evaluation of $\mathcal{A}$. Similarly to $readiness$, this action is either performed by $assert$ $\_strict(\mathcal{A})$ or $assert\_w\_agent(\mathcal{A})$. $s_a$ is hence defined as:
$$0\leq s_a(\mathcal{A}) \leq 1 =_{def} \left\{
\begin{array}{l}
	1 \text{ if } assert\_strict(\mathcal{A})=1\\
	1-2\sigma(agent_{assert}) \text{ otherwise }
\end{array}
\right.$$

Given these three measures, the consistency of a test case $tc=a_1\dots a_k \mathcal{A}_{k+1} \dots \mathcal{A}_{l}$ is defined as the mean of the consistency scores of each action:
$$0\leq consistency(tc) \leq 1 =_{def} 1/l(\sum_{1}^{k} (s_e(a_i)s_n(a_i)) + \sum_{k+1}^{l} s_a(\mathcal{A}_i))$$

\section{Test Case Unsoundness}
\label{sec:soundness}

Test case soundness is a critical property because it ensures that no conformant implementation is mistakenly classified as faulty. A test case is said unsound if there exists at least one conformant $AUT$ that is rejected by the test case \cite{Jard2000}. 

This section aims at delimiting the impact of agent use to execute NL test cases on soundness.

Soundness is typically defined with respect to formal models and requires a specification and an implementation relation.  However, NL test cases are not formal objects. 
In this context, a specification can be expressed with a deterministic input output automaton, e.g., an IOLTS $S=<Q, L \cup \T,\rightarrow,q_0>$, with $Q$ the state set, $L=L_I\cup L_O$ the set of inputs and the outputs, respectively, $\T$ the set of unobservable events, $\rightarrow$ the transition set, and $q_0$ the initial state. Informally, the traces of an IOLTS $S$, denoted $traces(S)$, are all the sequences of observable inputs and outputs obtained by covering the paths of $S$ (definitions in Annex). 

In our context, we suppose that there exists a specification $S$ whose inputs are navigation actions and outputs are GUIs. A NL test case $tc=a_1\dots a_k \mathcal{A}_{k+1} \dots \mathcal{A}_{l}$ is extracted from $S$ such that $\exists ?a_1 !g_1 \dots ?a_k !g_k \in traces(S)$ and $\mathcal{A}_{i} (k+1 \leq i \leq l)$ are true on the GUI $g_k$. With black-box testing, it is also assumed that $AUT$ can be expressed by an unknown IOLTS. This offers the advantage to define implementation relations that express the degree to which an implementation conforms to its formal specification. Several such relations could be considered. Here, we use $ioco$ \cite{treIoco96} as example, 
even though simpler relations such as trace equivalence could be applied. Under $ioco$, $AUT$ conforms to a specification $S$ if every action sequence derivable from $S$ produces, when executed on $AUT$, only outputs anticipated by $S$.

NL test cases are unsound for a plethora of reasons. For example, their actions are not formally defined as inputs or outputs, assertions may be ambiguous, etc. 
Algorithm \ref{algo:A1} dynamically builds the IOLTS $tc|AUT$ expressing the execution of $tc$ on $AUT$. This IOLTS adds the notions of inputs, outputs, along with internal actions that lead to terminal states labelled by verdicts. 

Despite the use of Algorithm \ref{algo:A1}, the execution of $tc$ is also unsound as the LLM agents $agent_{nav}$, $agent_{eval}$ and $agent_{assert}$ are used to interact with GUIs and to evaluate actions. These agents may hallucinate and return wrong responses. Let us assume that these agents have a probability of successfully performing their tasks that is close to 1. Despite this, $tc$ remains unsound because there still exists a small probability of producing false verdicts. In many practical contexts, however, this residual probability may be deemed acceptable when compared to other sources of uncertainty, such as environmental disturbances, which can similarly affect test execution.

From this observation, we introduce the notion of weak unsoundness, defined with respect to rare contexts in which a test case may be executed. In general terms, a test case is said weakly unsound if there exists a conformant implementation that can be rejected by the test case only under rare circumstances. 
We make that formal through the following definition. Let $\Omega$ denote the set of all possible contexts, and $\Omega_r$ denote a low-probability subset of $\Omega$ representing exceptional conditions. 

\begin{definition}[Weak Unsoundness]
	\label{def:weakunsound}	
	
Let $S$ be an IOLTS $<Q^S,L^S\cup \T,\rightarrow^S,q_0^S>$, $tc= a_1 \dots a_k \mathcal{A}_{k+1} \dots \mathcal{A}_l$ be a NL test case such that $\exists ?a_1!g_1 \dots ?a_k!g_k \in traces(S)$ and $\mathcal{A}_{j} (k+1\leq j\leq l)$ are true on $g_k$, and $tc|AUT$ be a deterministic IOLTS given by Algorithm \ref{algo:A1}.

\begin{itemize}
\item $AUT \text{ passes } tc$ iff  $\forall \sigma \in (L^S)^*, tc|AUT \overset{\sigma}{\nRightarrow} fail$
	
\item $tc$ is weakly unsound w.r.t. $S$  for $ioco$ iff $\forall AUT, AUT \text{ ioco } S$, $\forall c \in \Omega \setminus \Omega_r, AUT \text{ passes } tc \wedge \exists AUT, AUT \text{ ioco } S, \exists c \in \Omega_r,$ $\neg (AUT \text{ passes } tc)$
\end{itemize} 
\end{definition}

Within our context, weak unsoundness should reflect a practical tolerance for minor uncertainties inherent to the use of agents for the execution of NL test cases. Although such practical tolerance is difficult to quantify and depends on the application context, we propose a a quantitative characterisation in terms of sigma levels (Six Sigma (6$\sigma$) method \cite{6sigma}). Specifically, we require that agent performance goes above 3$\sigma$ ($p_3 = 0.9332$, $\sigma_3=0.2496$), knowing that 3$\sigma$ is the minimum acceptable level in many industries. The following proposition formalises this requirement.

\begin{proposition}
\label{prop:1}
Let $tc= a_1 \dots a_k \mathcal{A}_{k+1} \dots \mathcal{A}_l$ be a NL test case such that there exists $?a_1!g_1 \dots ?a_k!g_k \in traces(S)$, and $\mathcal{A}_{j} (k+1\leq j\leq l)$ are true on $g_k$, and $tc$ is made up of atomic actions and assertions. 
If $(\sigma(agent_{nav})<3\sigma$, $\sigma(agent_{eval})<3\sigma$, $\sigma(agent_{assert})<3\sigma)\in \Omega_r$, then $tc$ is weakly unsound.
\end{proposition}

Unfortunately, it is inherently difficult to state whether an action or assertion expressed in natural language is atomic and unambiguous. This issue can be addressed by exclusively writing NL test cases with actions and assertions that can be evaluated with $readiness_{strict}$ and $assert_{strict}$: 

\begin{proposition}
\label{prop:2}
If $tc$ is expressed only with navigation actions and assertions that can by evaluated with $readiness\_strict$ and $assert\_strict$, and $(\sigma(agent_{nav})<\sigma_3) \in \Omega_r$, then $tc$ is weakly unsound.
\end{proposition}

Proofs of these propositions are given in \cite{salva2025soundnessconsistencyllmagents}. 

Our experimental results indicate that, at present, Propositions \ref{prop:1} and \ref{prop:2}  hold with a few LLMs, which can be deployed on local servers. We believe that ongoing advancements in both LLM and agent technologies will soon render Propositions \ref{prop:1} and \ref{prop:2} attainable with more small language models.

\section{Experimental Results}
\label{sec:evaluation}

We investigated the capabilities of our algorithms through the following questions:

\begin{itemize}
	\item RQ1: How effectively can LLM agents perform navigation actions and evaluate readiness actions and assertions?
    
    \item RQ2: How do the estimated consistency measures $consistency(tc)$ compare with the observed test case execution consistency?
    
\end{itemize}

This study was conducted with 4 NL test suites:
\begin{itemize}
	\item TestG is a test suite designed for experimenting on 3 web sites (Google Gruyere, UCA, and a personal Web site). It includes 16 test cases, each consisting of 4 to 15 steps and 1 to 4 assertions. Each test case is supplemented with a boolean tabular specifying the expected outcome of every step. To broadly evaluate agent capabilities, the test cases include both positive assertions (where a Pass verdict is expected) and negative assertions (where a Fail verdict is expected). This enables the evaluation of our tool and agents under both correct and erroneous steps;
	
	\item TestA is a test suite specialised in assessing the capabilities of agents to evaluate assertions on 3 web sites (UCA, ARTEMIS, personal Web site). In our initial experiments, we observed that, occasionally, a navigation action deemed successful by the agent was in fact executed incorrectly, which in turn results in the failure of an assertion evaluation. Such inconsistencies hamper the accurate assessment of an agent's ability in evaluating assertions. To address this issue, we developed the test suite TestA, comprising NL test cases that contain only assertions. It includes 29 NL test cases, each having one assertion. As with TestG, a NL test case is associated with a boolean value to indicate the expected result of the assertion;  
	
	\item Test-W and Test-O are two test suites developed to experiment on two other distinct open-source e-commerce platforms: one for managing water related products \footnote{\url{https://github.com/dikshantnaik/Water-Management-System}} and the other one for managing electronic products \footnote{\url{https://github.com/opencart/opencart}}. Each NL test case consists of around 10 steps and one or two assertions. 
\end{itemize}

To perform these experiments, we developed 2 prototype tools composed of three agents. Two agents aim at evaluating readiness actions and assertions, as described in Section \ref{sec:tcexec}. The last agent was built on top of the Stagehand~\footnote{https://www.stagehand.dev/} framework, which automates browser interactions with natural language and code by means of LLMs. Stagehand was used to perform navigation actions and, in part, for data extraction from pages:  
\begin{itemize}
	\item EvalAgent is a tool specialised in computing, for a given agent, the 3 standard deviation scores $\sigma(agent_{readiness})$, $\sigma(agent_{nav})$, and $\sigma(agent_{assert})$, which assess the capabilities of an LLM agent to perform these three tasks; It takes a test suite and runs it $N$ times to compute standard deviations. The success of a navigation action is evaluated using the observe action;
	
	\item NLTestRunner implements Algorithm \ref{algo:A1}. Given a test suite and a number of runs $N$, it executes NL test cases, returns verdicts, and computes the estimated consistency measures and the observed test case execution consistencies if $N>0$. It can recognise 12 prompt forms to perform strict readiness and strict assertion evaluations.  
\end{itemize}
	
We performed our experiments with 8 LLMs, whose sizes range between 3B and 70B parameters. The LLM inference was carried out using Ollama\footnote{\url{https://ollama.com/}}, installed on a local server equipped with an NVIDIA L40 GPU (48 GB memory). All Web sites, test suites, and tools are available in \cite{companion25}.

\subsection{RQ1: How effectively can LLM agents perform navigation actions and evaluate readiness actions and assertions?}
\textbf{Setup:} 
To address this question, we considered these 8 LLMs: Qwen2.5:3B, Qwen2.5 7B, Mistral 7B, Mistral Nemo 12B, Deepseek-R1 14B, Qwen3:14B, Mistral Devstral 24B, and Llama3.1:70B. These models, provided by different companies, range from 3B to 70B parameters. We selected LLMs that support tool calling, which is a required feature to allow GUI interaction.

We evaluated them using the test suites TestG and TestA, executed with our tool EvalAgent. 

For each LLM used with the 3 agents of the tool EvalAgent, and for each NL test case, we evaluated the accuracy of each step category (readiness, navigation, and assertion) across 20 repeated runs. For each category, we computed both the (mean) accuracy and the standard deviation over these runs.

\begin{figure*}[htbp]
	\centering
	
	\begin{subfigure}[b]{0.47\textwidth}
		\centering
		\includegraphics[width=\textwidth]{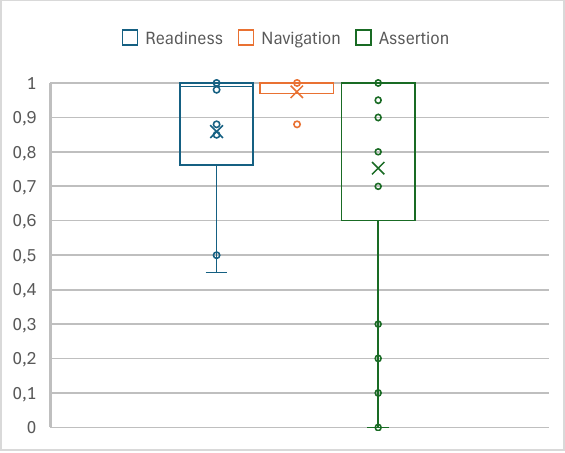}
		\caption{Qwen 2.5 3B}
		\label{fig:Qwen2.53B}
	\end{subfigure}
	\hfill
	\begin{subfigure}[b]{0.47\textwidth}
		\centering
		\includegraphics[width=\textwidth]{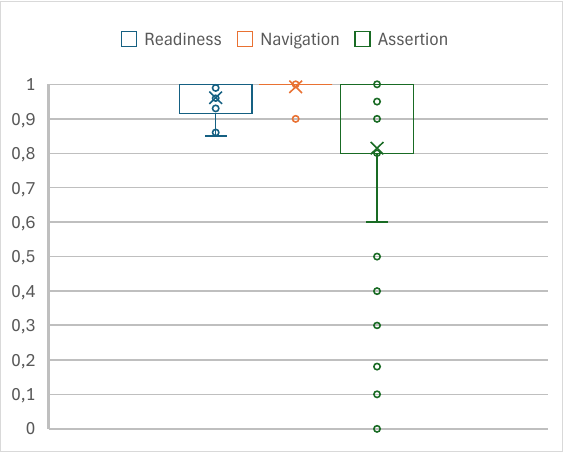}
		\caption{Qwen 2.5 7B}
		\label{fig:Qwen2.57B}
	\end{subfigure}
	
	\vskip\baselineskip
	
	\begin{subfigure}[b]{0.47\textwidth}
		\centering
		\includegraphics[width=\textwidth]{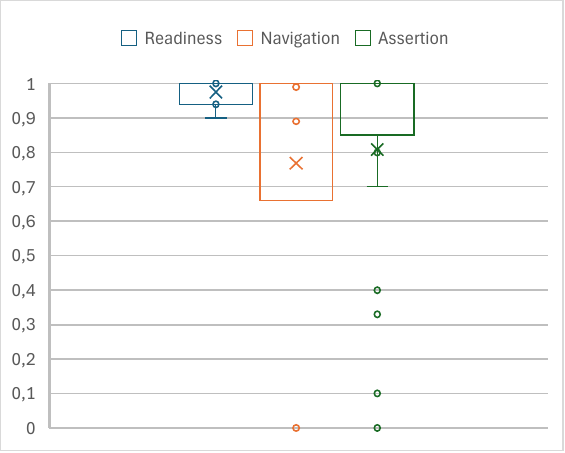}
		\caption{Mistral 7B}
		\label{fig:Mistral7B}
	\end{subfigure}
	\hfill
	\begin{subfigure}[b]{0.47\textwidth}
		\centering
		\includegraphics[width=\textwidth]{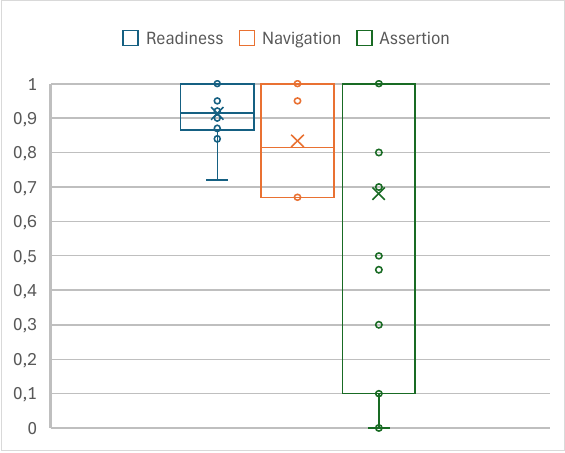}
		\caption{Mistral Nemo 12B}
		\label{fig:MistralNemo12B}
	\end{subfigure}
	
	
		\caption{Ability of LLM agent to perform readiness actions, navigation actions and assertions measured as mean accuracies with TestG and TestA over a batch of 20 runs (Part 1)}
	\label{fig:eval-llmcapabilities}
	\end{figure*}
	
	\begin{figure*}[htbp]
		\centering
		\begin{subfigure}[b]{0.47\textwidth}
			\centering
			\includegraphics[width=\textwidth]{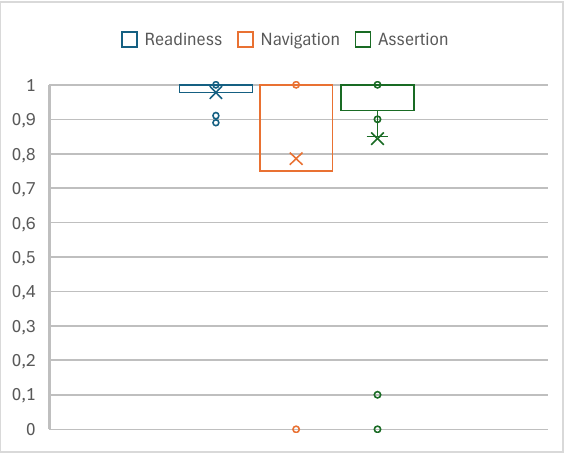}
			\caption{Qwen 3 14B}
			\label{fig:Qwen314B}
		\end{subfigure}
		\hfill
		\begin{subfigure}[b]{0.47\textwidth}
			\centering
			\includegraphics[width=\textwidth]{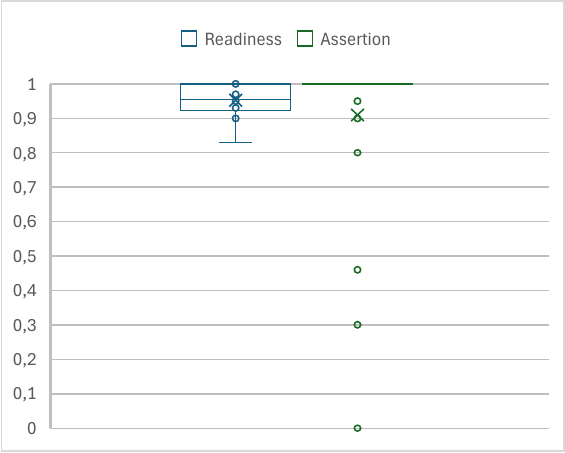}
			\caption{Deepseek R1 14B}
			\label{fig:DeepseekR114B}
		\end{subfigure}
		
		\vskip\baselineskip
		\begin{subfigure}[b]{0.47\textwidth}
		\centering	
	\includegraphics[width=\textwidth]{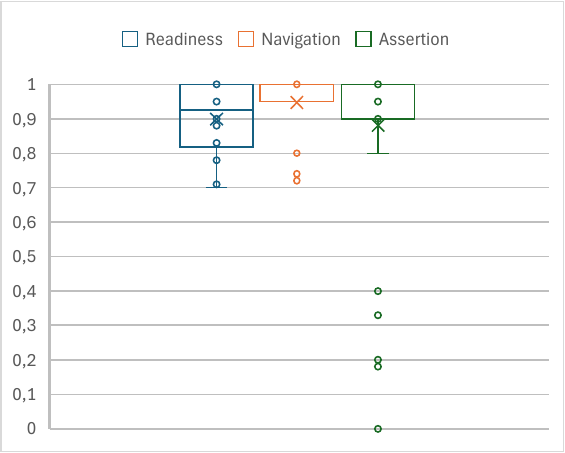}
		\caption{Mistral Devstral 24B}
		\label{fig:MistralDevstral24B}
	\end{subfigure}
	\hfill
	\begin{subfigure}[b]{0.47\textwidth}
		\centering
		\includegraphics[width=\textwidth]{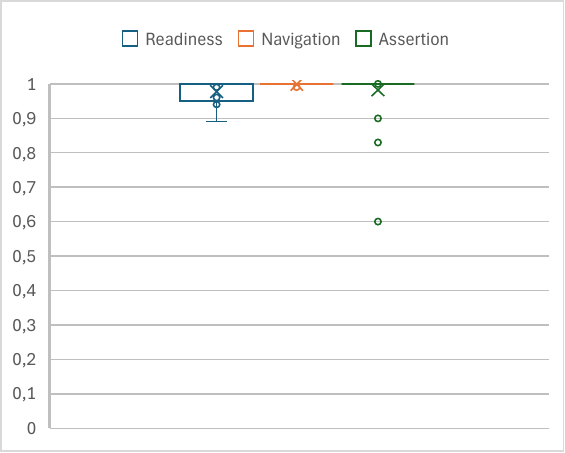}
		\caption{Llama 3.1 70B}
		\label{fig:Llama3.170B}
	\end{subfigure}
	
	\caption{Ability of LLM agent to perform readiness actions, navigation actions and assertions measured as mean accuracies with TestG and TestA over a batch of 20 runs (Part 2)}
	\label{fig:eval-llmcapabilities2}
\end{figure*}

\begin{table}[htbp]
	\centering
	\caption{Standard deviation results for readiness, navigation, and assertion actions.}
	\resizebox{.49\textwidth}{!}{
	\begin{tabular}{lccc}
		\toprule
	\textbf{LLM} & \makecell{\textbf{Readiness}\\\textbf{Std. dev.}} & \makecell{\textbf{Navigation}\\\textbf{Std. dev.}} & \makecell{\textbf{Assertion}\\\textbf{Std. dev.}} \\
	
		\midrule
		Qwen 2.5 3B           & 0.348& 0.158& 0.431\\ 
		Qwen 2.5 7B           & 0.192& 0.084& 0.388\\ 
		Mistral 7B            & 0.154& 0.421& 0.393\\ 
		Mistral Nemo 12B      & 0.280& 0.372& 0.466\\ 
		Qwen 3 14B            & 0.147& 0.410& 0.363\\ 
		Deepseek R1 14B       & 0.213& /    & 0.286\\ 
		Mistral Devstral 24B  & 0.300& 0.224& 0.324\\ 
		Llama 3.1 70B    & 0.149& 0.038& 0.132\\ 
		\bottomrule
	\end{tabular}
}
	\label{eval:stddev}
\end{table}

\textbf{Results:} Figures \ref{fig:eval-llmcapabilities} and \ref{fig:eval-llmcapabilities2} present box-plots 
of the distributions of mean accuracy across the NL test cases. The plots also show quartiles and mean values (indicated with crosses) of performing navigation actions, returning expected readiness, and returning expected assertion verdicts. Table \ref{eval:stddev} lists the standard deviations derived from these experiments.

Focusing on an easily interpretable measure, the mean accuracy, we observe a variability in performance across LLMs and tasks and identify three main groups of LLMs:

\begin{itemize}
\item Mean accuracies $>93.32$\% (level $3\sigma$): considering all test case step categories, only one model demonstrated acceptable capabilities in running NL test cases. Llama 3.1 70B achieved the best performance in our experiments, with mean accuracies greater than or equal to 98\%. However, the box plot reveals some outliers in assertion evaluations;

\item Mean accuracies $\geq 80$\%: Qwen 2.5 7B, DeepSeek R1, and Devstral 24B fall into this group. The box plot indicates that Qwen 2.5 7B has good capabilities in executing readiness and navigation actions, but produces mixed results when evaluating assertions. DeepSeek R1 is unable to execute navigation actions despite being a model that supports tool calling. It shows good capabilities in evaluating readiness and assertions but it is less effective than Llama 3.1 70B. 

\item Mean Accuracies $<80$\%: the remaining four models Qwen 2.5 3B, Mistral 7B, Mistral Nemo 12B, and Qwen 3 14B can execute some NL test cases correctly, but they frequently fail to execute navigation actions. Among them, Qwen 3 14B is close to the threshold, with a mean accuracy of 78\% for navigation, 98\% for readiness, and 84\% for assertions. While it succeeds on many navigation tasks, it struggles more than Llama 3 70B, particularly on certain websites (e.g., the personal website).  
Qwen 2.5 3B seems to be effective for GUI interaction but performs poorly in evaluating assertions. This reveals a sort of bias in our experiments. We indeed observed that failed assertion evaluations may sometimes come from navigation actions that the LLM agent does not execute correctly. Some LLM agent can indeed return a success in performing a navigation action despite an incorrect real execution. Since the GUI state then diverges from the expected one, the assertions fail. Our current NL test case dataset, TestG, does not allow precise detection of navigation errors. This is why we also use TestA in these experiments. 
\end{itemize}

For the last two performance groups, we analysed the testing traces and identified three primary reasons why the LLMs may have failed to perform the actions as expected:
\begin{itemize}
\item their context lengths are sometimes too short (the number of available tokens for LLM input and output) to consider all the page content;

\item the LLM ability to extract page content is variable. Our code and prompts for performing this extraction do not take into account the structured representation of the data, such as tabular forms, which the assertion evaluation relies on. Furthermore, our prompts could be more suited to some LLMs than others.

\item Some NL test steps may be interpreted as ambiguous by certain LLMs and not by others. For instance, the assertion ``Assert that the term 'Course' is present 4 times" is interpreted differently: some LLMs expect exactly four occurrences, while others interpret it as at least four. Additionally, the same LLM may have different interpretations of the same assertion across several runs. 
\end{itemize}

Initial conclusions: 1) Considering these results together with Table \ref{eval:stddev}, we can conclude that only Llama 3.1 70B demonstrates good capabilities in NL test case execution while maintaining high consistency (standard deviation  $<0.15$). However, there remains room for improvement to achieve higher performance. 2) Evaluating LLM capabilities in performing navigation actions would benefit from more complex datasets. For example, designing test suites of NL test cases, in which each navigation step is associated with an expected GUI screenshot could yield more precise measurements of accuracy and standard deviation. 3) Our tools currently face technical constraints, such as limited LLM context size and data extraction challenges. These issues, however, can be addressed with improvements in tooling and prompt design. 4) NL test cases may include ambiguous steps. Our approach requires an additional step to identify and manage such ambiguities. 5) None of the LLMs used in our experiments are specialised for test case execution. LLMs specifically trained or fine-tuned for GUI interaction could provide better results. 

\subsection{RQ2: How do estimated consistency measures $consistency(tc)$ compare with the observed test case execution consistency?}

\textbf{Setup:} In RQ1, we measured the mean accuracies and standard deviations for several LLM agents. Based on these results, we selected three representative models: Llama 3 70B, as the most capable LLM in executing NL test cases; Mistral Nemo 12B, as the least capable; and Qwen3 14B, which demonstrated mixed performance.

Given a NL test case $tc$ and the previously obtained results, particularly the standard deviations of an LLM agent, we can now employ NLTestRunner, which implements Algorithm~\ref{algo:A1}, to execute $tc$ and compute the estimated consistency measure, $consistency(tc)$. Furthermore, by executing $tc$ in batches of 20 runs, we also obtained an empirical measure of consistency of the returned test verdict. We applied our tool with the test suites TestG, Test-M, and Test-O. 

To compare the observed execution consistency of NL test cases with our measure $consistency(tc)$, we calculate the mean relative error (MRE) for each test case. MRE is a standard statistical measure used to quantify how far predictions deviate from reference values. 

\begin{figure*}[htbp]
	\centering
	\begin{subfigure}[b]{0.32\textwidth}
		\centering
		\includegraphics[width=\textwidth]{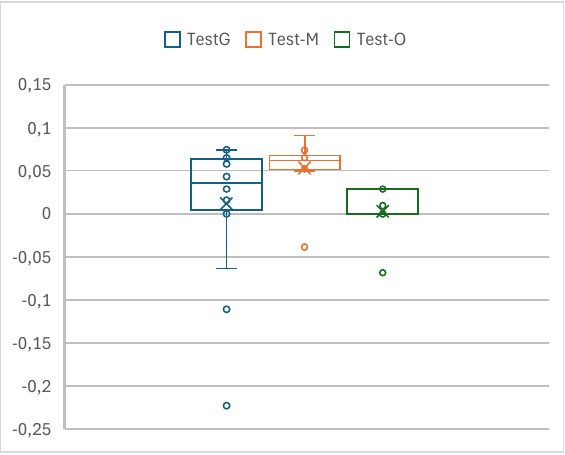}
		\caption{Qwen 3 14B}
		\label{fig:sub1}
	\end{subfigure}
	\hfill
	\begin{subfigure}[b]{0.32\textwidth}
		\centering
		\includegraphics[width=\textwidth]{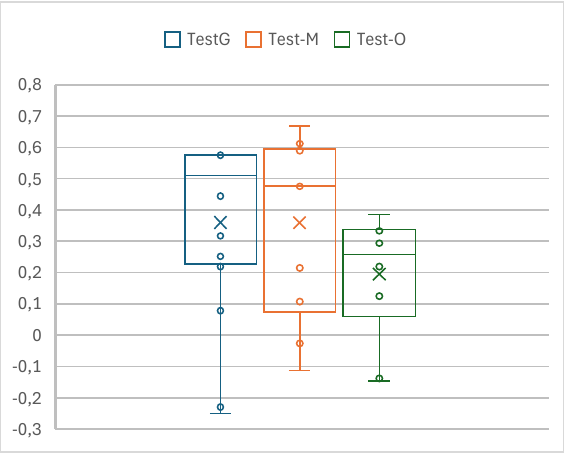}
		\caption{Mistral Nemo 12B}
		\label{fig:sub2}
	\end{subfigure}
	\hfill
	\begin{subfigure}[b]{0.32\textwidth}
		\centering
		\includegraphics[width=\textwidth]{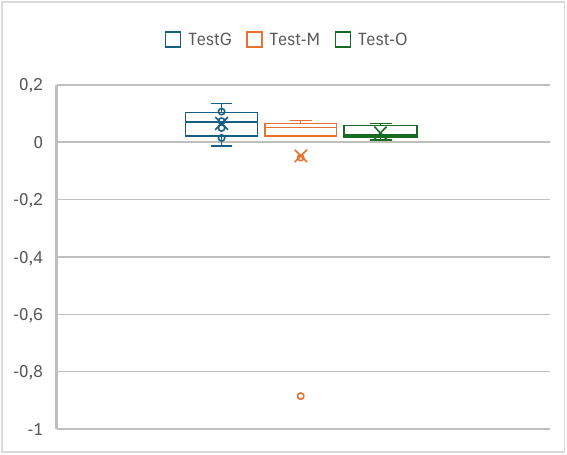}
		\caption{Llama 3.1 70B}
		\label{fig:sub3}
	\end{subfigure}
	\vskip\baselineskip
\caption{Box plots depicting the MRE measured used to quantify how far estimated consistencies deviate from real consistencies with the test suites with TestG, Test-M and Test-O.}
\label{fig:eval-consistency}
\end{figure*}
\textbf{Results:} For each LLM, Figure \ref{fig:eval-consistency} presents 3 box plots that show the distributions of MRE computed after the execution of each test case, for each test suite.  The MRE means are depicted with crosses.

Aggregating across all test suites and LLMs, we obtain a mean MRE of 11\%. This indicates that, on average, our estimated measure deviates from the observed stability of NL test case executions by a non-negligible margin. However, Figure \ref{fig:eval-consistency} shows a pronounced difference in MRE between the models. Mistral Nemo 12B shows the highest mean MRE of 30\%, whereas Llama 3.1 70B and Qwen 3 14B achieve much lower values, with a mean MRE of 2\%. 

For Mistral Nemo 12B, $consistency(tc)$ exhibits substantial deviations from the observed execution consistencies of the test cases. Specifically, in one-third of the cases where the LLM agent produces the expected results, and in half of the cases where it does not, $consistency(tc)$ underestimates the observed consistency. This indicates that when the ability of the LLM agent is limited, our measure $consistency(tc)$ becomes inaccurate. In contrast, for LLM agents with mixed or strong capabilities in executing NL test cases, $consistency(tc)$ produces estimates that closely align with the observed consistencies. 

It is worth noting that, in the case of Qwen 3 14B, a high consistency score does not necessarily imply correct behaviour. Rather, it indicates that the model tends to behave in a highly predictable manner, either producing the expected outcome very frequently, or failing to do so with the same consistency.

Initial conclusions: 1) our experiments indicate that $consistency(tc)$ is an accurate measure for estimating the execution consistency of a test case when the LLM has mixed or good abilities in NL test case executions. 2) For other LLMs, however, a more fine-grained metric, potentially combining multiple measures for different types of assertions or GUI interactions, could provide estimates that more closely align with the observed execution consistency.

\subsection{Threats To Validity}
\label{threats} 
We now address potential threats to the validity of our evaluation. We organise them into internal and external threats.

\textbf{Internal threats.} 
The first set of threats concerns factors that may have affected our experimental results. These include limitations in our ptototype tools, the design of NL test cases, the writing of prompts, and the extraction of data.
1) The evaluation of LLM agents depends on our tools, which could be improved to better support response timeouts, capture structured data from GUIs, and verify whether navigation actions are executed correctly. 
2) The NL test suites we developed are made up of common interactions but do not cover all cases, such as hovering over or dragging elements
3) While we took care to avoid writing ambiguous NL test steps, ambiguity is a difficult property to judge. Some works considered this problem \cite{userstory-ambiguity}, but further work is needed.   
4) The design of the prompts used in our experiments may also limit the validity of our conclusions, particularly when applied to other LLMs. Despite following a strict protocol and applying prompts consistently across models, they may still be better suited to some LLMs than to others.

\textbf{External threats.} 
The second set of threats concerns factors that may affect the generalisability of our findings. External threats include the choice of the $AUT$, the selection of LLMs, and the use of Stagehand to perform navigations.

1) The $AUT$s we considered in our experiments consist of only 5 different web sites. Furthermore, while we included some negative assertions in the test suites, the $AUT$s themselves are not supposed faulty. To strengthen external validity, further kinds of applications, such as mobile applications, should be considered.  
2) We considered 8 LLMs that can be deployed on local servers, but excluded cloud-based LLMs that may provide better results for GUI interaction. We only focused in this paper on LLMs that can be deployed on local servers as they offer the advantage of ensuring confidentiality. These LLMs come from different companies, with model sizes ranging from 3B to 70B parameters. However, future work could explore the evaluation of our tools with further LLMs. 
3) We used the framework Stagehand for performing navigation actions, which is relatively new and still evolving. New versions of Stagehand may yield better results, especially for small LLMs (e.g., 7B LLMs). Other frameworks could also be considered. For example, we conducted some experiments with BrowserUse, which is a framework rather dedicated to the execution of general scenarios. However, our initial results were not promising.

\section{Conclusion}
\label{sec:conclusion}

In this paper, we investigated the feasibility of executing NL test cases for GUI applications using LLM agents. To support this study, we introduced formal notions of weak unsoundness and proposed the use of Six-Sigma levels to characterise the abilities of LLM agents. In addition, we defined a complementary measure to quantify the consistency of NL test case execution. Building on this, we developed an algorithm with guardrail mechanisms and specialised agents that dynamically verify the test step executions. To evaluate our approach, we designed several NL test suites and implemented prototype tools to execute them. Experiments across 8 LLMS of varying sizes showed that current LLMs can handle GUI testing, under some conditions: no ambiguity in actions and assertions, and LLMs being able to perform navigation and evaluation tasks with high accuracy. As we demonstrated, Meta Llama 3.1 70B surpasses the 3-Sigma threshold in NL test case execution with a high level of execution consistency, while smaller LLMs struggled with navigation or assertion evaluation. 

This work also opens several future directions. 
Can NL test cases be generated from higher-level scenarios or navigation maps? How can our tools be improved to optimise GUI data extraction, handle response timeouts, and support more complex assertion structures?
Can screenshots be included into test sets to better evaluate agent capabilities? 
How can our algorithm be extended to balance properties like soundness, laxness, controllability, and efficiency? Can NL test cases be designed to evaluate non-functional aspects, such as security? Can small LLMs be fine-tuned to specialise in some specific tasks, while larger ones only handle complex tasks? 
If agent compositions could be optimised dynamically to GUI complexity, this might provide more efficient executions. 

\section{Acknowledgement}
Research supported by the Industrial Chair on Reliable and Confident Use of LLMs (https://uca-fondation.fr/les-chaires/) and MIAI Cluster, France 2030 (ANR-23-IACL-0006)
\bibliographystyle{apalike}
{\bibliography{doc} }

\appendix
\section{Annex}
\label{sec:annex}
\subsection{Prompts Used By Our Agents}

\begin{figure}[htbp]
	\centering
	\begin{tcolorbox}[title=Navigation Prompt, width=0.49\textwidth, colback=white!95!gray, colframe=black, sharp corners=south]
		\begin{Verbatim}[fontsize=\scriptsize]
You are an autonomous agent whose role is to perform 
interactions on Web pages with a web browser.
			
# ROLE AND OBJECTIVE
You will be given an instruction that describes an 
interaction to be performed on the web page. Your task 
is to execute the corresponding interaction on the page.
			
# PROCESS
1. Identify the element in the instruction 
2. Find the exact corresponding element in the page.
3. Use a function to interact with the page
4. Generate a set of facts that are contained in the output
			
# Common functions:
click, fill, type, press, or any other playwright locator 
method
			
# OUTPUT FORMAT
You must respond with valid JSON strictly using the 
following format: 
{{ "facts": ["fact 1", "fact 2", "..."], "task_accomplished"
		: "Success|Failed|Unknown" }}
		\end{Verbatim}
	\end{tcolorbox}
	\caption{Scenario to search for ARTEMIS project news}
	\label{fig:navprompt}
\end{figure}

\begin{figure}[htbp]
	\centering
	\begin{tcolorbox}[title=Readiness Evaluation Prompt, width=0.49\textwidth, colback=white!95!gray, colframe=black, sharp corners=south]
		\begin{Verbatim}[fontsize=\scriptsize]
You are an evaluation agent tasked with 
determining whether an action can be performed 
on a Web page.

#ROLE AND OBJECTIVE
You will be given a page content and an action.
Your task is to check if an action can be 
performed on the page. 
The page content is a list of elements
formatted as {{id, description, type'}}
Read the descriptions and the types of the 
elements carefully
Respond 'True' if the action can be performed
on the page and 'False' otherwise.
Let think step by step and return the final
response.

# PROCESS
1. Identify all the elements related to the 
action (link, statictext, etc.) in the given 
page content 
2. Extract the descriptions and types of the 
elements
3. Check if the interaction given in the 
action is possible
3. Conclude if the action can be performed

#OUTPUT FORMAT
You must respond with valid JSON strictly 
using the following format: 
{{ "facts": ["fact 1", "fact 2", "..."], 
		"result": true|false }}		
		\end{Verbatim}
	\end{tcolorbox}
	\caption{Scenario to search for ARTEMIS project news}
	\label{fig:evalprompt}
\end{figure}

\begin{figure}[htbp]
	\centering
	\begin{tcolorbox}[title=Assertion Evaluation Prompt, width=0.49\textwidth, colback=white!95!gray, colframe=black, sharp corners=south]
		\begin{Verbatim}[fontsize=\scriptsize]
You are an evaluation agent tasked to evaluate 
an Assertion of a test case.
			
#ROLE AND OBJECTIVE
You will be given a page content and an 
assertion.
Your task is to evaluate the assertion on the 
page content. 
The page content is a list of elements 
formatted as {{id, description, type'}}
Read the descriptions and the types of the 
elements carefully
Respond 'True' if the Assertion is True and 
'False' otherwise
Let think step by step and return the final 
response.
			
# PROCESS
1. Identify all the elements related to the 
assertion (link, statictext, etc.) in the 
given page content 
2. Extract the descriptions and types of the 
elements
2. Check if some elements meet the assertion 
3. Conclude on the result of the assertion 
based on your observations

#OUTPUT FORMAT
You must respond with valid JSON strictly
 using the following format: 
{{ "facts": ["fact 1", "fact 2", "..."], 
		"Verdict": true|false }}		
		\end{Verbatim}
	\end{tcolorbox}
	\caption{Scenario to search for ARTEMIS project news}
	\label{fig:assertprompt}
\end{figure}

\subsection{NL Test Case Soundness}
We use the classical following notations of IOLTS:

\begin{definition}
Let $S=<Q,L\cup \T, \rightarrow, q_0> $be an IOLTS and $\mu_i \in L_I\cup L_O \cup \T$,  $a_i \in L_I \cup L_O, q, q',q_{i} \in Q$.

\begin{itemize}
\item $q \xrightarrow{\mu_1...\mu_n} q' =_{def} \exists q_0, q_n : q=q_0 \xrightarrow{\mu_1} q_1 \dots \xrightarrow{\mu_n} q_n=q'$
\item $q \xRightarrow{\epsilon} q' =_{def} q=q' \vee q \xrightarrow{\tau_1...\tau_n} q'$
\item $q \xRightarrow{a} q' =_{def} \exists q_1, q_2 : q \xRightarrow{\epsilon} q_1 \xrightarrow{a} q_2 \xRightarrow{\epsilon} q'$
\item $q \xRightarrow{a_1...a_n} q' =_{def} \exists q_0, q_n : q=q_0 \xRightarrow{a_1}q_1 \dots \xRightarrow{a_n} q_n=q'$

\item $S \xRightarrow{\sigma} q' =_{def} \exists q': q_0 \xRightarrow{\sigma} q'$
\item $traces(q) =_{def} \{ \sigma \in L^* \mid \exists q' :q \xRightarrow{\sigma} q' \}$

\item $traces_\T(q) =_{def} \{ \sigma \in  (L \cup \T )^* \mid \exists q' :q \xrightarrow{\sigma} q' \}$
\item $q \text{ after } \sigma = \{q' \mid q \xRightarrow{\sigma} q'\}$
\item $out(q) =_{def} \{ a \in L_O \mid \exists q' \in Q: q\xRightarrow{a}\}$. $out(Q) =_{def} \bigcup_{q\in Q} out(q)$
\end{itemize}
\end{definition}

\begin{definition}[$ioco$ implementation relation]
Let $AUT$ and $S$ be two IOLTS. $AUT\text{ }ioco\text{ }S \Leftrightarrow_{def} \forall \sigma \in traces(S): out(AUT \text{ after } \sigma ) \subseteq out(S \text{ after } \sigma)$
\end{definition}	

We rewrite Definition \ref{def:weakunsound} with these notations (adapted from the definitions given in \cite{Jard2000}):

\begin{definition}[Weak Unsoundness]
	\label{def:weakunsound2}		
Let $S=<Q^S,L^S\cup \T,\rightarrow^S,q_0^S>$ be a deterministic IOLTS and $tc= a_1 \dots a_k \mathcal{A}_{k+1} \dots \mathcal{A}_l$ be a NL test case such that $\exists ?a_1!g_1 \dots ?a_k!g_k \in traces(S)$ and $\mathcal{A}_{j} (k+1\leq j\leq l)$ are true on $g_k$. 
	
$tc$ is weakly unsound w.r.t. $S$  for $ioco$ iff\\
$\forall AUT, AUT \text{ ioco } S, \forall c \in \Omega \setminus \Omega_r, \forall \sigma \in traces(S) \cap traces(AUT) \cap traces(tc|AUT), \forall !g \in L_O^S$ such that $\sigma.!g\in traces(tc), !g \in out(AUT \\ \text{ after } \sigma), !g \in out(S \text{ after } \sigma), tc \text{ after } \sigma!g \neq FAIL$ $\wedge \exists AUT, AUT$ $\text{ ioco } S, \exists c \in \Omega_r, \exists \sigma \in traces(S) \cap traces(AUT) \cap traces(tc|AUT),$ $\exists !g \in L_O^S$ such that $\sigma.!g\in traces(tc), !g \in out(AUT \text{ after } \sigma), !g \in out(S \text{ after } \sigma),$ $tc|AUT \text{ after } \sigma!g = FAIL$.  
\end{definition}

\subsection*{Proofs Proposition \ref{prop:1} and \ref{prop:2}}

Let $\sigma= ?a_1!g_1 \dots ?a_i (i> 0) \in traces(S)$ and $!g \in L_O^S$ . The actions $readiness$, $?a_i$, $\mathcal{A}_{j}$ can be performed by the agents $agent_{eval}$, $agent_{nav}$, and $agent_{assert}$, having the following behaviours: A) the agents are not faulty (correct responses, no hallucination, no errors) B) the agents return errors; C) the agents hallucinate (wrong responses) under rare circumstances, i.e., under the context $c \in \Omega_r$. 

We now list all the traces and verdicts that can be obtained with $\sigma !g |AUT$ w.r.t. A) B) and C):

A) $traces_\T(\sigma !g|AUT)$ in \[
\left\{
\begin{array}{l}
	\sigma !g \\
	\sigma !g \text{ observe }\\
	\sigma !g \text{ observe }\text{ readiness }\\
	\sigma !g \text{ observe }\text{ readiness } \mathcal{A}_{k+1} \dots \mathcal{A}_{l}, \text{ if } g=g_k \text{ since }\mathcal{A}_{k+1} \dots \mathcal{A}_{l}\\ \quad\text{true on }g_k
\end{array}
\right. 
\]
Hence, $tc|AUT \text{ after } \sigma !g \neq FAIL$. $tc$ is sound. 

B) $traces_\T(\sigma !g|AUT)$ in \[
\left\{
\begin{array}{l}
	\sigma !g\\
	\sigma !g \neg \text{observe } \text{(error by $agent_{nav}$, no GUI}\\ \quad \text{update)}\\
	\sigma !g \text{ observe } \neg\text{readiness } \text{(error by }agent_{nav},\\ \quad \text{wrong GUI that cannot accept the}\\\quad \text{next action)}\\
	\sigma !g \text{ observe } \neg\text{readiness } \text{(error by } agent_{readiness})\\
	\sigma !g \text{ error } \text{(error returned by $agent_{assert}$)}\\
	\sigma !g \text{ observe } \text{readiness } \text{ (use of $readiness\_strict$)}\\
	\sigma !g \text{ observe }\text{ readiness } \mathcal{A}_{k+1} \dots \mathcal{A}_{l} \text{ (use of}\\\quad assert\_strict) 
\end{array}
\right.
\]
Hence, $tc|AUT \text{ after } \sigma !g \neq FAIL$. $tc$ is sound.

C) Firstly, let us assume that $agent_{nav}$ operates always correctly, but both $agent_{readiness}$ and $agent_{assert}$ return incorrect results.

$traces_\T(\sigma !g|AUT)$ in \[
\left\{
\begin{array}{l}
	\sigma !g\\
	\sigma !g \text{ observe }\\
	\sigma !g \text{ observe } \neg\text{readiness } \text{ ($agent_{readiness}$ returns}\\ \quad \text{ wrong response)}\\
	\sigma !g \neg\mathcal{A}_{k+1} \text{ ($agent_{assert}$ fails)}\\
	\sigma !g \text{ observe } \text{readiness } \text{(use of $readiness\_strict$)}\\
	\sigma !g \text{ observe }\text{readiness } \mathcal{A}_{k+1} \dots \mathcal{A}_{l} \text{ (use of}\\\quad assert\_strict)
\end{array}
\right.
\]

Hence, $tc|AUT \text{ after } \sigma !g = FAIL$ in 4th trace. $tc$ is unsound. If we suppose that $agent_{assert}$ behaves incorrectly under rare circumstances (context $c \in \Omega_r$), then Proposition \ref{prop:1} holds. If We suppose that $readiness\_strict$ and $assert\_strict$ are used instead of $agent_{readiness}$ and $agent_{assert}$, then 3rd and 4th traces do not belong to $traces_\T(\sigma !g|AUT)$ anymore. $tc|AUT \text{ after } \sigma !g \neq FAIL$. $tc$ is sound. 

Now, let us assume that $agent_{nav}$ does operate correctly, it returns a wrong GUI $!g_f \neq g$. $traces_\T(\sigma !g|AUT)$ in \[
\left\{
\begin{array}{l}
	\sigma !g_f\\
	\sigma !g_f \neg \text{ observe} \text{ (wrong gui returned by $agent_{nav}$)}\\
	\sigma !g_f \text{ observe } \text{readiness } \text{ (wrong response by $agent_{readiness}$)}\\
	\sigma !g_f \text{ observe } \neg\text{readiness } \text{ (use of $readiness\_strict$)}\\
	\sigma !g_f \mathcal{A}_{k+1} \dots \mathcal{A}_{l} \text{ ((wrong assertion evaluations by $agent_{assert}$) }\\
	\sigma !g_f \mathcal{A}_{k+1} \dots \neg \mathcal{A}_{k+j} \text{ (wrong assertion evaluations ended with}\\ \text{       a correct evaluation assertion by $agent_{assert}$) }\\
	\sigma !g_f \neg \mathcal{A}_{k+1} \text{ (use of $assert\_strict$)}\\
\end{array}
\right.
\]

$tc|AUT \text{ after } \sigma !g = FAIL$ in 6th and 7th traces. $tc|AUT$ is unsound even with the assumption that $assert\_strict$ is used. If we suppose that both $agent_{assert}$ and $agent_{nav}$ behave incorrectly under rare circumstances (context $c \in \Omega_r$), then $tc$ is weakly unsound, and Proposition \ref{prop:1} holds. If only $agent_{nav}$ behaves incorrectly under rare circumstances and if 
$readiness\_strict$ and $assert\_strict$ are used instead of $agent_{readiness}$ and $agent_{assert}$, then 3rd, 5th and 6th traces do not belong to $traces_\T(\sigma !g|AUT)$ anymore. $tc$ is weakly unsound, and Proposition \ref{prop:2} holds. 

It is worth noting that in traces 3 and 5, the wrong behaviour of $agent_{nav}$ is not detected. It is masked by the wrong behaviour of $agent_{reasdiness}$ or $agent_{assert}$. 
  
\end{document}